# Electrical tuning of exciton-plasmon polariton coupling in monolayer MoS$_2$ integrated with plasmonic nanoantenna lattice

**Bumsu Lee**[1†], **Wenjing Liu**[1†], **Carl H. Naylor**[2], **Joohee Park**[1], **Stephanie Malek**[1], **Jacob Berger**[1], **A.T. Charlie Johnson**[2] **and Ritesh Agarwal**[1*]

[1]Department of Materials Science and Engineering, University of Pennsylvania, Philadelphia, PA 19104, USA

[2]Department of Physics and Astronomy, University of Pennsylvania, Philadelphia, PA 19104, USA

[†] These authors contributed equally to this work.

[*] To whom correspondence should be addressed. E-mail: riteshag@seas.upenn.edu

**Active control of light-matter interactions in semiconductors is critical for realizing next generation optoelectronic devices, with tunable control of the system's optical properties via external fields. The ability to manipulate optical interactions in active materials coupled to cavities via geometrical parameters, which are fixed along with dynamic control with applied fields opens up possibilities of controlling exciton lifetimes, oscillator strengths and their relaxation properties. Here, we demonstrate electrical control of exciton-plasmon polariton coupling strength of a two-dimensional semiconductor integrated with plasmonic nanoresonators assembled in a field-effect transistor device between strong and weak coupling limits by electrostatic doping. As a result, the exciton-**

**plasmon polarion dispersion can be altered dynamically with applied electric field by modulating the excitonic properties of monolayer MoS₂ arising from many-body effects with carrier concentration. In addition, strong coupling between charged excitons plasmons was also observed upon increased carrier injection. The ability to dynamically control the optical properties of an ultra-thin semiconductor with plasmonic nanoresonators and electric fields demonstrates the versatility of the coupled system and offers a new platform for the design of optoelectronic devices with precisely tailored responses.**

Optical and electronic properties of mono and few-layered two-dimensional (2D) transition metal dichalcogenide (TMDC) semiconductors have been receiving great attention due to their very unique attributes originating from strong two-dimensional confinement and inadequate exciton screening along with new potential applications including ultra-thin optoelectronic devices, exciton-polariton condensation at high temperature, and spin-valley devices (*1-3*). The strongly confined 2D excitons in monolayer TMDC semiconductors lead to several interesting optical features such as very large exciton binding energy (*4-7*) and oscillator strength (*8, 9*), and the observation of higher-order excitations such as charged excitons (trions) (*10-11*). The optical properties of 2D excitons can be significantly altered when coupled to an external cavity, leading to enhancement of spontaneous emission rate and lasing in the weak exciton-photon coupling regime, and the formation of exciton-polaritons, in the strong coupling regime (*12-20*). Various types of optical cavities including distributed Bragg Reflector (DBR) (*12, 13*), photonic crystal (*14*) and plasmonic nanoresonators (*15-21*) have been utilized to manipulate the optical properties of 2D excitons demonstrating weak, intermediate and strong coupling regimes. Of the different cavity types, plasmonic nanoantenna arrays offer unique opportunities for tailoring light-matter interactions in 2D materials owing to their ease of integration and open geometry, which is also compatible with electrical injection or gating device configuration. Furthermore, plasmonic lattices depending on the lattice geometry allows for tunable and intense resonances originating from localized surface plasmons (LSPs) coupled to the lattice dispersion known as "lattice-LSP" modes (*22*, *23*). The lattice-LSPs are propagating modes with strong electric fields, and relatively high quality factors in comparison to LSP resonances from individual (uncoupled) nanoparticles. Recently, monolayer molybdenum disulfide ($MoS_2$) integrated with silver nanoantenna array has shown large enhancements of the Raman scattering and photoluminescence in the weak coupling regime (*15-20*), Fano resonances in the intermediate coupling range (*16*) and normal mode splitting (vacuum Rabi splitting) with the formation of exciton-plasmon polaritons in the strong coupling regime (*21*).

On the other hand, owning to the atomically thin nature of the monolayer TMDC semiconductors and the associated significantly reduced screening of the charged carriers, their optoelectronic properties can be controlled via electrostatic doping induced by the gate electric field in a field-effect transistor (FET) geometry. As a result, novel findings have been reported including the modulation of exciton binding energies (*24*, *25*), observation of negatively charged trions (*10-11*), and valley-contrast second-harmonic generation (*26*). Here, by combining a plasmonic lattice cavity coupled with monolayer $MoS_2$ configured in a FET device, we demonstrate active control of the coupling strengths between the $MoS_2$ excitons and lattice-LSP resonance, with continuous and reversible transition between the strong and weak exciton-plasmon coupling regimes achieved by electron injection or depletion in the active $MoS_2$ monolayer. Furthermore, under certain doping conditions, strong coupling between trions and plasmons was also observed, which demonstrates the level of active control of light-matter coupling that can be achieved in these 2D systems.

To fabricate the optoelectronic device (Fig. 1A), monolayer MoS$_2$ was grown directly on Si/SiO$_2$ substrate by chemical vapor deposition (*27*) (Supplementary Methods), followed by the fabrication of source and drain electrodes and silver plasmonic nanodisk array (Supplementary Methods). A 285 nm-thick thermally grown SiO$_2$ layer sandwiched between MoS$_2$ and highly doped Si substrate produces a typical metal-oxide-semiconductor FET with a capacitance of ~12 nF/cm$^2$. Charges are injected or depleted from the electrical contacts depending on the gate electrostatic potential ($V_G$) with respect to MoS$_2$. For example, a $V_G$ of ~100 V can lead to a carrier density of ~8×10$^{12}$/cm$^2$ in the MoS$_2$ monolayer, estimated from the capacitance of the device. As a result, the chemical potential (Fermi level) of MoS$_2$ can be altered by ~50 meV depending on the carrier density. Figure 1B shows the transconductance curve obtained from a bare MoS$_2$ FET device, displaying a typical *n*-type channel performance. To fabricate the plasmonic lattice cavity, the diameters of the silver nanodisks were varied between 90 and 140 nm, with a constant thicknesses and pitch of 50 and 460 nm respectively, to tune the LSP energies near the MoS$_2$ excitonic region to enable the lattice-LSP dispersions to strongly couple to excitons to form polaritons (*21*). To characterize the evolution of the coupling strength between the MoS$_2$ excitons and lattice-LSPs, the dispersions of the coupled MoS$_2$ exciton and lattice-LSP system were measured by a home-built angle-resolved reflection microscopy setup (*28*) while tuning the gate voltage (Supplementary Methods).

The angle-resolved differential reflectance spectra (ΔR/R = (R$_{sample}$-R$_{substrate}$)/R$_{substrate}$), for MoS$_2$ monolayer integrated with silver nanodisk array with disk diameters, $d$ = 100 nm, measured at 77 K shows anti-crossing and kink-like behavior near the MoS$_2$ A- and B-exciton energies (yellow dashed lines in Fig. 1D), respectively, indicating strong coupling between the MoS$_2$ excitons (flat dispersion) and lattice-LSPs of (±1, 0) diffractive order (Fig. 1F) at $V_G$ = 0; When a gate bias was applied, significant dispersion changes in the angle-resolved spectra were observed (Fig. 1C-F). Upon sweeping the $V_G$ from 0 to -60 V, the MoS$_2$ exciton-plasmon system evolves progressively to display stronger exciton-lattice LSP coupling, indicated by a clear anti-crossing behavior in dispersion curves, as a result of carrier depletion. On the contrary, as the $V_G$ was ramped up to +60 V, the measured dispersion of the system gradually became linear with the MoS$_2$ exciton dispersion crossing the lattice-LSP dispersion, implying that the strong exciton-lattice LSP coupling was gradually changed to weak coupling upon carrier injection.

Interestingly, we found strong coupling of the lattice-LSP modes to both the neutral excitons (A$^o$) and charged trions (A$^-$), leading to the formation of their own discrete polariton branches (Fig. 1G). We will focus our attention to spectral changes in the vicinity of the A$^-$ exciton resonance for simplicity due to its relatively larger exciton oscillator strength compared to B-excitons (*5*, *10*). Uncoupled A$^o$ and A$^-$ states measured from the bare sample region are located at approximately 646 nm and 653 nm, respectively (Supplementary Information). Figures 1G shows a series of line-cuts obtained at different $V_G$, extracted from the angle-resolved spectra of the array at an angle where the lattice-LSP mode has zero detuning with the A$^o$ exciton (sin θ = 0.21, vertical white-dashed line in Fig. 1D as an example). For a clear illustration, ΔR/R spectra show peaks corresponding to both A$^o$ and A$^-$ polariton states clearly resolved in the 620 – 660 nm wavelength range with their evolution indicated by grey-dashed lines. As $V_G$ decreases to -100 V from 0 V, the coupling of A$^o$ exciton and lattice-LSP became stronger with an increased polariton mode splitting owing to charge carrier depletion. In contrast, upon increasing $V_G$, the A$^o$ polariton modes gradually lose intensity and get merged in the reflection spectra, indicating that the system evolves from strong to weak coupling regime. On the other hand, the A⁻ polariton mode (~654 nm) gains oscillator strength, resulting in more intense reflection peaks and polariton mode splitting, which suggests that the system changed progressively from strong $A^o$ exciton-plasmon coupling to strong A⁻ exciton-plasmon coupling regime with increasing carrier concentration.

In order to characterize the exciton-plasmon coupling tuning phenomena in $MoS_2$ FET devices in detail, we designed a plasmonic lattice with LSP resonance very close to the A-excitons by changing the silver nanodisk diameters to $d$ = 120 nm (Fig 2). Angle-resolved spectra were fitted to a coupled oscillator model (COM) in order to obtain exciton-plasmon coupling strengths (Supplementary Methods) of the system as it evolves with $V_G$. The exciton-plasmon coupling was reflected by a blue shift of these polariton modes with respect to the uncoupled $A^o$ and A⁻ states measured from the bare sample region, which resulted from the effective repulsive interaction with LSP resonances (Fig. 2A-F, blue-dashed lines) located at the red side of the excitonic region. The coupling between the lattice-LSP modes with both exciton and trion was observed to be modulated by $V_G$. At $V_G$ = 0 V, only the $A^o$ exciton-polariton appeared in the reflectance spectrum of $MoS_2$ with strong coupling to the lattice-LSP modes, as indicated by a clear anti-crossing behavior of the dispersion curves (Fig. 2A) with a coupling strength of 57 meV. The strong $A^o$ exciton-plasmon coupling and the absence of A⁻ exciton polarion features indicate that the sample is lightly doped (exciton-plasmon coupling strength at $V_G$ = 0 slightly varies depending on the as-grown sample's initial doping state). As the carrier concentration increases upon applying positive $V_G$, the $A^o$ exciton displays a slight blue shift in its resonance energy (yellow dashed lines in Fig. 2A-F) and linewidth broadening in the ΔR/R spectra due to many-body effects upon carrier doping (*24*) (Supplementary Information), and was gradually decoupled from the lattice-LSP modes to eventually disappear in the dispersion spectra (Figs. 2A-F). The gate voltage-dependent exciton-LSP coupling strength calculated by the COM (Fig. 2G) shows that at $V_G$ >80 V, the $A^o$ exciton coupling to lattice-SP modes vanishes, owing to the collapse of its oscillator strength under high doping conditions, thus representing a large change in coupling strength from 57 meV to almost zero as a function of positive $V_G$. The strong coupling between trions and lattice-LSP modes was also observed upon increased doping. At intermediate $V_G$ (Fig. 2C), the observed polariton modes originate from both the $A^-$ (lower energy) and $A^o$ (higher energy) excitons, as indicated by gray arrows. In contrast to $A^o$ excitons (Fig. 2G), the coupling of trions to plasmons exhibited an opposite trend with increased electron doping, reaching a maximum value of 35 meV at $V_G > 60$ V from zero at no applied gate bias. Lower value in the maximum (saturated) coupling strength of $A^-$ exciton can be explained by its weaker oscillator strength than $A^o$ exciton. However, given that the typical oscillator strength of trions found in 2D quantum wells or bulk materials is more an order of magnitude lower than that of the neutral excitons (*10, 29*), this finding is intriguing since the magnitude of the coupling strength of trions with plamsoncs is comparable to $A^o$ exciton-plasmon coupling, which further illustrates the enhanced interaction between excitons and charged carriers due to significantly reduced screening in monolayer semiconductors. In control experiments (Supplementary Information), reflectivity spectra of bare monolayer $MoS_2$ configured as an FET device also reveals decrease (increase) of oscillator strengths of the $A^o$ ($A^-$) exciton transitions upon electrostatic doping and only a very small shift of their resonance energies due to many-body effects (*24, 25*). These results are consistent with our observation of the evolution of the exciton-plasmon polariton modes and their coupling strengths with increased doping, as will be discussed later in more detail.

To further study the modulation of light-matter coupling of the system and its active control via external fields, nanodisk arrays with different disk diameters ($d =$110 and 140 nm) were prepared on the same large area $MoS_2$ monolayer. LSP resonances arising from different sized silver disks resulted in different coupling strengths with $A^o$ and $A^-$ excitons (Fig. 3) due to different LSP-exciton detuning (*21*). Silver nanodisks with diameters of $d = 110$ and 140 nm have their LSP resonances at 648 nm and 685 nm (Figs. 3A-C and E-G, blue dashed lines) respectively. The mode at 648 nm is in resonance with the $A^-$ exciton energy, while the 685 nm mode is largely detuned from the excitonic region (>150 meV detuning). The effective tuning of the mode dispersions (Fig. 3A-C and E-G) with different $V_G$ was observed in both devices, with similar trends (Fig. 3D and H) as observed for the device in Fig. 2. However, the exciton-plasmon coupling strengths of the array with LSP in resonance with the excitons ($d = 110$ nm) were significantly higher than the array with detuned LSP resonance ($d = 140$ nm) at the same $V_G$. These results, consistent with other similar devices, also provide more insights about design strategies for obtaining precisely tailored optoelectronic responses in 2D polaritonic devices.

Increased many-body effects are one of the unique characteristics of highly confined TMDC monolayers even at moderate doping levels, which can be easily achieved from $SiO_2$/Si back-gated FET devices (*7*, *10*, *24*, *25*). Greatly reduced screening due to the lack of large bulk polarization leads to increased Coulomb interaction and scattering process, which along with phase space-filling (Pauli blocking) effect influences the exciton oscillator strength and binding energy (*29*, *30*). With increasing charge carrier concentration with positive electrostatic gating, reduction of the $A^o$ oscillator strength due to exciton bleaching from these many-body effects gives rise to weaker interactions between excitons and plasmons, driving the system towards weak coupling. On the contrary, the depletion of charge carriers with negative gate potential renders MoS$_2$ more intrinsic, resulting in stronger exciton-plasmon coupling. Thus, continuous and reversible switching between strong and weak exciton-plasmon coupling can be achieved in response to the variation in charge carrier concentration (Fermi level shift) in the TMDC monolayer, which can be easily controlled by external fields, and is a unique attribute of a 2D system.

In summary, we demonstrated active control of the coupling strength between 2D MoS$_2$ excitons interacting with plasmonic lattices integrated in an FET device, via charge carrier injection/depletion in MoS$_2$. Continuous and reversible transition between the strong and weak exciton-plasmon coupling regime was achieved by tuning the polarity and magnitude of the gate voltage. Furthermore, we also showed that the trion resonances can also couple strongly with the lattice plasmons at high electron doping concentration, forming trion-plasmon polaritons. Our work demonstrates that unique opportunities exist for 2D monolayer semiconductors owing to the lack of bulk polarization that reduces carrier screening leading to extraordinarily strong exciton oscillator strengths and binding energies, and increased many body effects, thereby enabling excellent control over exciton-polariton properties via external fields. Electrical control of light-matter interactions in photonic devices is very attractive for designing modulators, switches and sensors and for active tuning of polariton dispersion in functioning devices.

**Methods**

**Device fabrication:** Monolayer MoS$_2$ flaks were grown on SiO$_2$/Si substrate (280 nm thick SiO$_2$ layer) via chemical vapor deposition method. Silver nanodisk array and the source and drain electrodes were patterned directly on the as-grown MoS$_2$ substrate by electron beam lithography, followed by the deposition of 50 nm silver by electron-beam evaporation. More detailed information for both MoS$_2$ growth and fabrication of the plasmonic nanostructures and electrical contacts have been described earlier (*16*, *21*, *27*). The sample was transferred to a chip carrier with electrical pins designed for our optical microscopy cryostat (Janis ST-500) and the sample electrodes are connected to electrical pads of the chip carrier by wire-bonding for the electrical measurements.

**Optical and electrical measurements.** For angle-resolved reflection spectra, the measurement setup and the detailed information about the measurements are described elsewhere in our previous study (*28*). For the electrical measurements, two Keithley 2400 were used to apply the source-drain bias and the gate voltage, and the channel current was recorded through Keithley 5617 electrometer.

**Coupling constant calculation by coupled oscillator model (COM):** Five oscillators (A-exciton, trion, lattice-LSP modes with (±1, 0) diffractive orders, LSPs) were used in the COM to fit our angle-resolved reflectance data of MoS$_2$ coupled with plasmonic lattice as shown in Figs. 2 and 3. The Hamiltonian as per the COM is expressed as,

$$H = \begin{bmatrix} E_A - i\gamma_A & 0 & g_{A-L} & g_{A-L} & g_{A-LSP} \\ 0 & E_T - i\gamma_T & g_{T-L} & g_{T-L} & g_{T-LSP} \\ g_{A-L} & g_{T-L} & E_{L+} - i\gamma_{L+} & 0 & g_{L-LSP} \\ g_{A-L} & g_{T-L} & 0 & E_{L-} - i\gamma_{L-} & g_{L-LSP} \\ g_{A-LSP} & g_{T-LSP} & g_{L-LSP} & g_{L-LSP} & E_{LSP} - i\gamma_{LSP} \end{bmatrix}$$

where E and γ denote the transition energy and the damping (half-width at half-maximum) of each mode, respectively. The subscripts *A*, *T*, *L+*, *L-* , and LSP stand for A-exciton, trion, (+1,0), (-1,0) diffractive orders of lattice-LSP modes, and LSP resonance, respectively. The off-diagonal terms are the coupling strengths between each coupled resonance while the diagonal terms represent the uncoupled states. The coupling between the A-exciton and trions, and between the two diffractive modes were ignored (set to zero), but other coupling strengths were used as fitting parameters. For more detailed information, please refer to (*21*).

**Figure Captions:**

**Fig. 1**. **Electrical tuning of exciton-plasmon polariton coupling in monolayer MoS$_2$ coupled with plasmonic nanodisk array configured in a FET device.** (**A**), Schematic diagram of the MoS$_2$ monolayer with plasmonic lattice integrated in an FET device geometry for carrier injection and depletion. (**B**), Transconductance data obtained from bare MoS$_2$ FET device. Inset: SEM image of the MoS2-plasmonic lattice device. (**C** to **F**), Angle-resolved differential reflectivity (ΔR/R) contour plots measured at various gate voltages ($V_G$). A- and B-excitons are indicated as yellow-dashed lines in (D). (**G**) ΔR/R spectra obtained from line cuts (sin$\theta$ = 0.21, vertical white-dashed line in (D)) of the angle-resolved spectra at different $V_G$. Evolution of A$^o$ and A$^-$ polariton states are shown by grey-dashed lines. The ΔR/R spectra are offsetted for clarity.

**Fig. 2**. **Evolution of the exciton-plasmon coupling between both neutral (A$^o$) and charged (A$^-$) excitons, and lattice-LSP modes under different electron doping conditions.** (**A** to **F**). Angle-resolved ΔR/R contour plots and the corresponding coupled oscillator model (COM) fits are presented at different gate voltages. At $V_G$ = 40 V, A$^o$ and A$^-$ polariton modes are indicated in (C). The position of A$^o$ and A$^-$ excitons, and LSP resonances are indicated as yellow, green and blue-dashed lines, respectively in (A to F). (**G**) Calculated Coupling strengths for A$^o$ and A$^-$ exciton-plasmon polariton modes as obtained from the COM are plotted as a function of the gate voltage.

**Fig. 3**. **Gate-voltage dependent exciton-plasmon coupling in the angle-resolved ΔR/R plots for two different silver nanodisks arrays with different resonance detuning values.** (**A** to **C**) $d = 110$ nm. (**E** to **G**) $d = 140$ nm. The position of $A^o$ and $A^-$ excitons, and LSP resonances are indicated as yellow, green and blue-dashed lines, respectively. (**D**) and (**I**), Comparison of the exciton-plasmon coupling strengths for $A^o$ and $A^-$ polariton modes as obtained from the COM as a function of the gate voltage.

**Figures**

**Figure 1.**

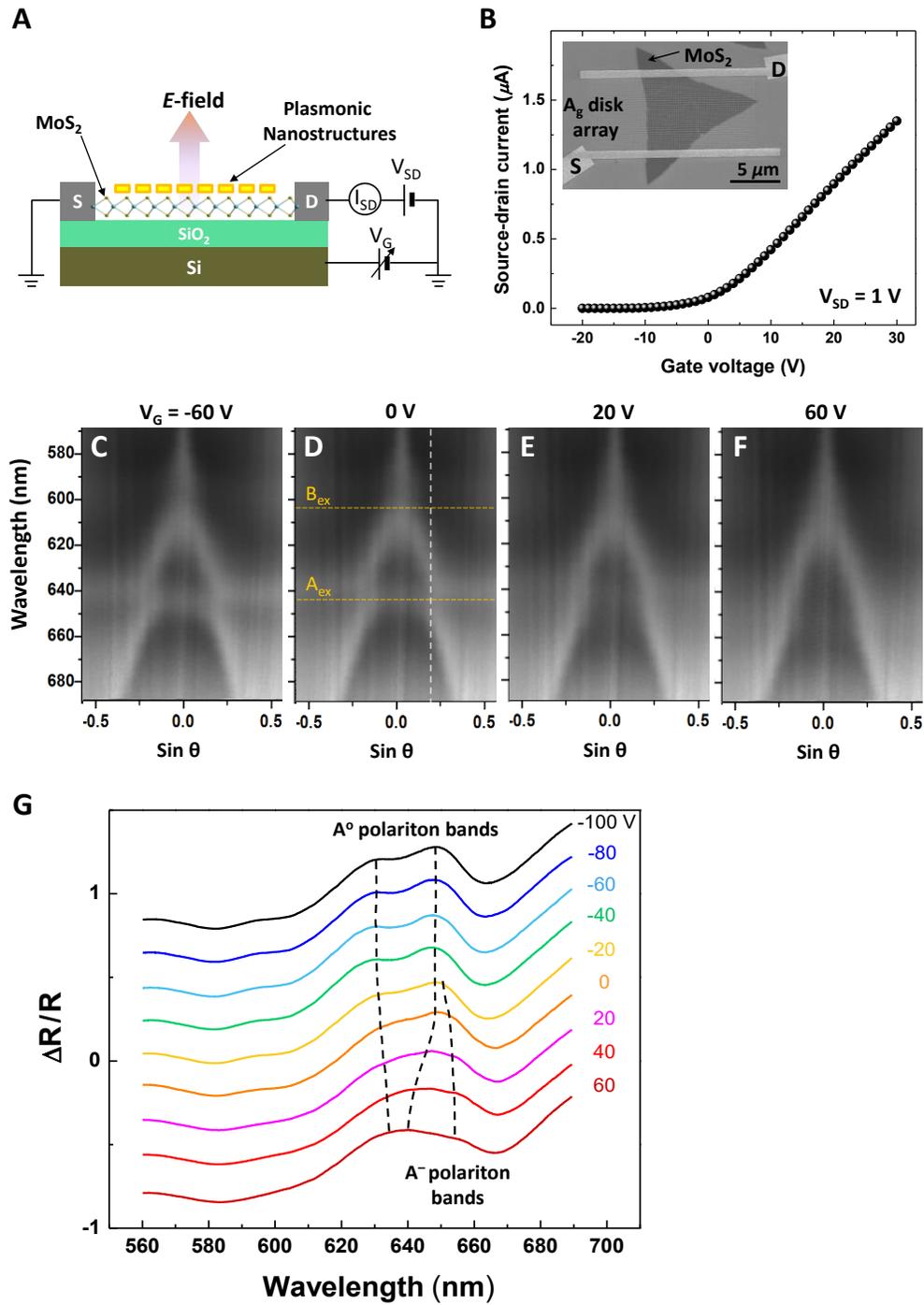

**Figure 2.**

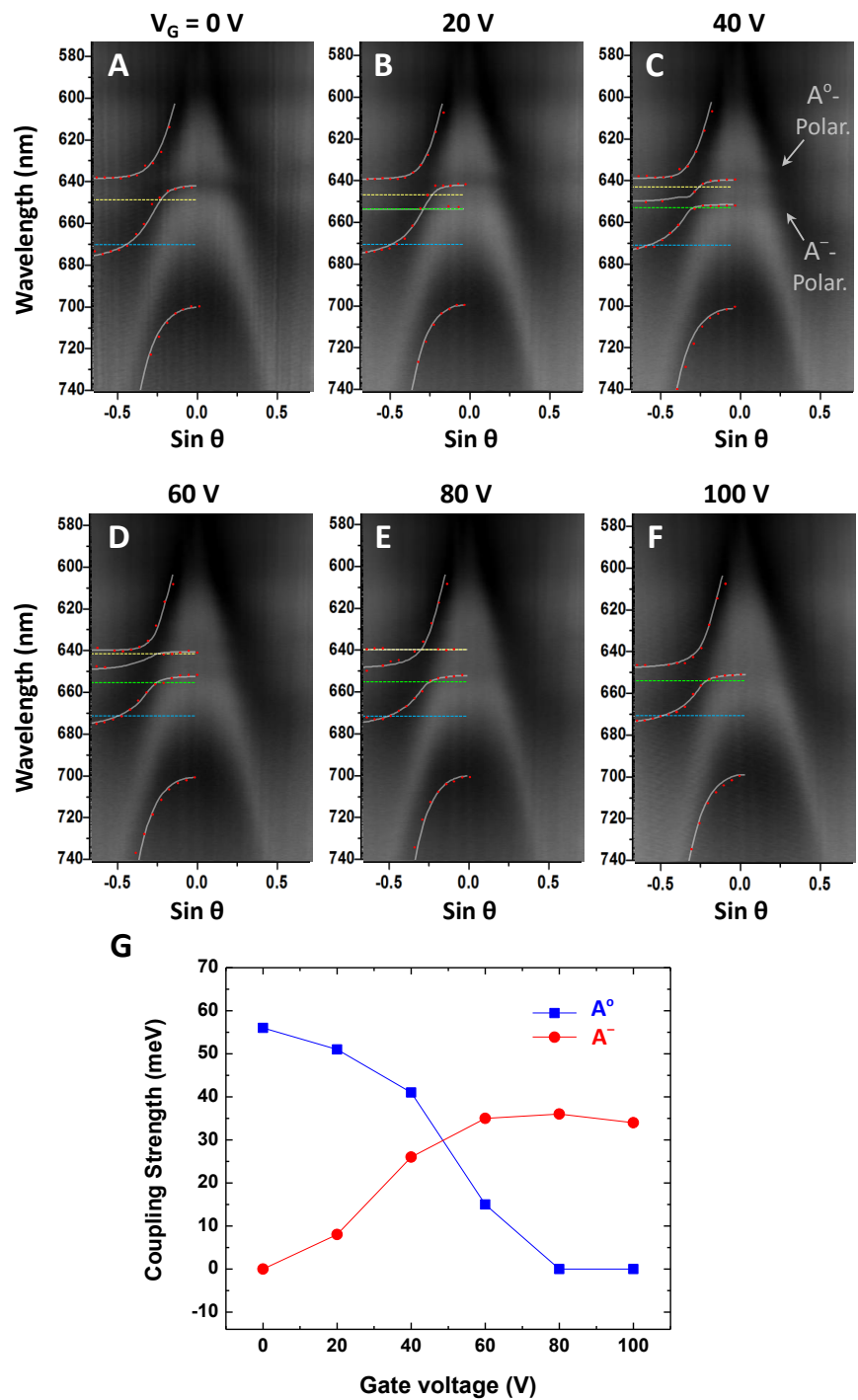

**Figure 3.**

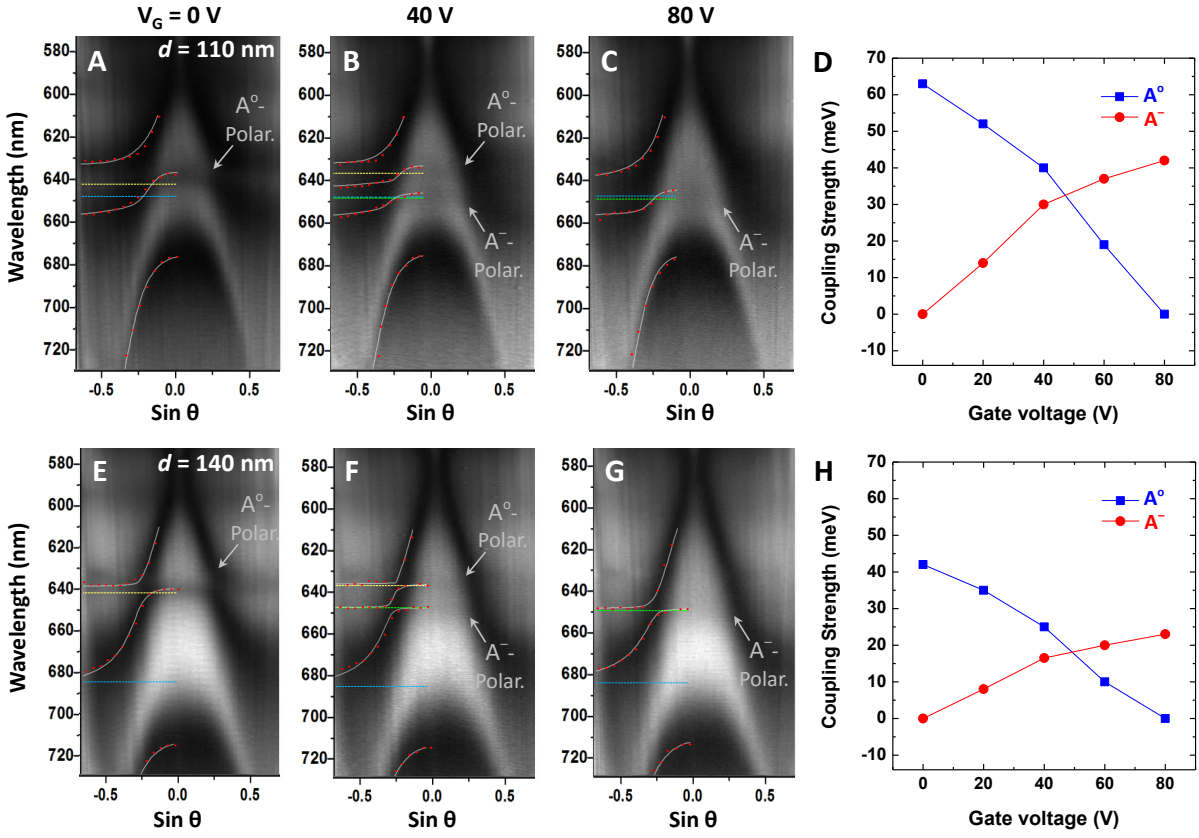